\documentclass[prl,nofootinbib,twocolumn]{revtex4}

\usepackage{amssymb}
\usepackage{bm}
\usepackage{amsmath}

\usepackage{graphicx}
\usepackage{fleqn}
\usepackage{longtable}
\usepackage{verbatim}
\usepackage{amsfonts}

\usepackage{float}

\newcommand{\be}{\begin{equation}}
\newcommand{\ee}{\end{equation}}
\newcommand{\bea}{\begin{eqnarray}}
\newcommand{\eea}{\end{eqnarray}}
\newcommand{\beq}{\begin{equation}}
\newcommand{\eeq}{\end{equation}}
\def\beqa{\begin{eqnarray}}
  \def\eeqa{\end{eqnarray}}
\newcommand{\bv}{\left(\begin{array}{c}}
\newcommand{\ev}{\end{array}\right)}

\def\lsim{\mathrel{\rlap{\lower4pt\hbox{\hskip1pt$\sim$}}
    \raise1pt\hbox{$<$}}}         %less than or approx. symbol
\def\gsim{\mathrel{\rlap{\lower4pt\hbox{\hskip1pt$\sim$}}
    \raise1pt\hbox{$>$}}}         %greater than or approx. symbol

\newcommand{\bra}[1]{\left\langle{#1}\right\vert}
\newcommand{\ket}[1]{\left\vert{#1}\right\rangle}

\begin{document}

\begin{flushright}
DO-TH 12/10 
\end{flushright}

\vspace*{-30mm}

\title{\boldmath Extracting $B \to K^*$ Form Factors from Data}

\author{Christian Hambrock}
\email{christian.hambrock@udo.edu}
\author{Gudrun Hiller}
\email{gudrun.hiller@udo.edu}
\affiliation{Institut f\"ur Physik, Technische Universit\"at Dortmund, D-44221 
Dortmund, Germany}

\vspace*{1cm}

\begin{abstract}
	We extract ratios of $B \to K^*$ form factors  at low hadronic recoil from recent data on $B \to K^* \mu^+ \mu^-$ decays  in a model-independent way.
The presented method
will improve in the future with
	further  (angular) studies in semileptonic rare $B$-decays and  advance our understanding of form factors, which are important inputs  in precision tests of the Standard Model.
	\end{abstract}

\maketitle

%%%%%%%%%%%%%%%%%%%%
{\bf Introduction.}
The experimental studies of  flavor changing neutral current (FCNCs) decays of $b$-flavored hadrons are  advancing rapidly with the successful start of the Large Hadron Colliders' (LHC) $b$-physics program.  This generates further demand for good control of hadronic
uncertainties to enable precision tests of the Standard Model (SM).

A  leading source of
theoretical uncertainty in semileptonic FCNC
decays $B \to M \bar \ell \ell$, $\ell=e,\mu,\tau,\nu$ of a $B$ meson into a light meson $M$
are the transition form factors.
When the energy of the emitted meson is large in the $B$ meson rest frame, the method of light cone QCD sum rules (LCSR) applies and allows to calculate the form factors in this kinematical regime \cite{Ball:1998kk}. 
The region in which the emitted meson is softly recoiling against the decaying $B$ meson is accessible to lattice QCD. Preliminary unquenched results are available \cite{Liu:2011raa}.
As we rely on such predictions to probe electroweak and flavor physics, it is
important to obtain further independent information.

The low recoil region 
allows for observables in which the
 form factors can be probed model-independently~\cite{Bobeth:2010wg}. 
Two of these short-distance free observables, the fraction of longitudinally polarized vector mesons $F_L$ and the transverse asymmetry $A_T^{(2)}$~\cite{Kruger:2005ep},
have recently been measured by CDF~\cite{Aaltonen:2011ja}, BaBar~\cite{BaBarLakeLouise2012} ($F_L$ only) and LHCb~\cite{LHCbMoriond2012}  in the decay $B \to K^* \mu^+ \mu^-$.
Here, we extract ratios of $B \to K^*$ form factors  from  these data. Since the ratios  are obtained at low recoil they can be directly compared to lattice 
predictions. 
 It is the aim of this letter to demonstrate the performance of this new model-independent method from the first time available precision data.

{\bf $\bm{B \to K^* \ell^+ \ell^-}$ at low recoil.}
The transversity amplitudes $A_{\perp,||,0}^{L,R}$ in $B \to K^* \ell^+ \ell^-$ decays at low recoil 
factorize at lowest order \cite{Grinstein:2004vb} and share universal short-distance coefficients $C^{L,R}$  \cite{Bobeth:2010wg}
\begin{equation} \label{eq:benefit}
A_{i}^{L,R} \propto C^{L,R} f_i , ~~~~~~~~~i=\perp,||,0\, ,
\end{equation}
while the form factors $f_i$ are independent of the Wilson coefficients of the electroweak theory.
 A fourth transversity amplitude
  arises for finite lepton mass $m_\ell$. However, at low recoil it is suppressed by
 $m_\ell^2/m_B^2$, where $m_B$ denotes mass of the  $B$-meson, and can be safely neglected.
(We suppress the dependence on the dilepton invariant mass squared $q^2$ for brevity in this work.)

It follows that observables of type $(A^L_i A^{L*}_j \pm A^R_i A^{R*}_j)/(A^L_l A^{L*}_k \pm A^R_l A^{R*}_k )$, where
$i,j,k,l=\perp,||,0$, probe long-distance physics only. This includes
$F_L$ and $A_T^{(2)}$, given as  \cite{Kruger:2005ep}
\begin{align}
  F_{L} & =
  \frac{|A_0^L|^2 + |A_0^R|^2}{\sum_{X=L,R} ( |A_0^X|^2+ |A_\perp^X|^2+|A_\parallel^X|^2 )}\,,  
  \label{eq:FL} \\
    A_T^{(2)} & =
  \frac{|A_\perp^L|^2 + |A_\perp^R|^2-|A_\parallel^L|^2-|A_\parallel^R|^2}
       {|A_\perp^L|^2 + |A_\perp^R|^2+|A_\parallel^L|^2+|A_\parallel^R|^2}\, .
\end{align}
Using Eq.~(\ref{eq:benefit}) one obtains at fixed $q^2$~\cite{Bobeth:2010wg},
\begin{align}    \label{eq:fitform}
  F_{ L} & = 
  \frac{f_0^2}{f_0^2 + f_\perp^2 + f_\parallel^2},
  \quad                   
A_T^{(2)}  = \frac{f_\perp^2 - f_\parallel^2}{f_\perp^2 + f_\parallel^2} .
  \end{align}
 The $f_i$  read 
   \begin{align}\label{eq:ffdef}
   	   f_\perp & =  {\cal{N}}  \frac{\sqrt{2 \hat s \hat{\lambda}}}{1 + \hat m_{K^*}} V, \quad
  f_\parallel
=   {\cal{N}}   \sqrt{2 \hat s}\, (1 + \hat m_{K^*})\, A_1, 
  \nonumber
\\  
  f_0 & =  {\cal{N}} 
    \frac{(1 - \hat{s} - \hat m_{K^*}^2) (1 + \hat m_{K^*})^2 A_1 - \hat{\lambda}\, A_2}
    {2\, \hat m_{K^*} (1 + \hat m_{K^*}) } , 
\end{align}
with
\begin{align}\label{eq:factordef}
	{\cal{N}} &=G_F \alpha_e  V_{tb} V_{ts}^* \sqrt{
\frac{ m_B^3  \sqrt{\hat \lambda}}{3\times 2^{10}\pi^5}
}
,~
\hat s
=
\frac{q^2}{m_B^2}
,~
\hat m_{K^*} = \frac{m_{K^*}}{m_B},
\nonumber\\
\hat \lambda
&=
1+\hat s^2 +\hat m_{K^*}^4- 2(\hat s + \hat s \hat m_{K^*}^2+\hat m_{K^*}^2).
\end{align}
The relevant $B \to K^*$ $q^2$-dependent form factors $V,A_1$ and $A_2$ are defined as usual~\cite{Ball:2004rg}: 
\begin{align}
	&\bra{K^*(k, \epsilon)} \bar{s} \gamma_\mu b \ket{B(p)}\label{eq:ffbasisBZfirst}
=
\frac{2 \, V }{m_B + m_{K^*}} \varepsilon_{\mu\rho\sigma\tau} \epsilon^{*\rho} p^\sigma k^\tau ,
        \\
  &\bra{K^*(k, \epsilon)} \bar{s} \gamma_\mu \gamma_5 b \ket{B(p)}
         = 
\nonumber\\
&i\epsilon^{*\rho} 
            \left[ 2 A_0\,m_{K^*}   \frac{q_\mu q_\rho}{q^2} + A_1 \,(m_B + m_{K^*})  (g_{\mu\rho} - \frac{q_\mu q_\rho}{q^2})  \right.  \nonumber
  \\
   &   - \left.A_2\, q_\rho \left(
   \frac{(p+k)_\mu}{m_B + m_{K^*}} - \frac{m_B - m_{K^*}}{q^2} (p-k)_\mu \right)\right] ,
   \label{eq:ffbasis2}
\end{align} 
where $\epsilon$ denotes the polarization vector of the $K^*$.

It is clear from Eqs.~(\ref{eq:fitform}) and \eqref{eq:ffdef}, that one can obtain ratios of form factors
$V/A_1$ and $A_1/A_2$ by resorting to the observables $F_L$ and $A_T^{(2)}$. 
We comment on the limitations of this determination:
The universality of  Eq.~(\ref{eq:benefit}) is broken by power corrections at order $1/m_b$ to 
the improved Isgur-Wise form factor relations between
 the form factors of the dipole currents $T_{1,2,3}$ \cite{Ball:2004rg}
 and the form factors of the axial and axial-vector currents.
 However, these corrections enter the transversity amplitudes with additional parametric suppression by the Wilson coefficients $C_j$ of the $|\Delta B|=|\Delta S|=1$ effective Hamiltonian
 $C_7/C_9 \sim {\cal{O}}(0.1)$  \cite{Grinstein:2004vb,Bobeth:2010wg}. Hence, we expect corrections at most at the level of a few percent, as for the  $1/m_b$-corrections to the 
 operator product expansion (OPE)~\cite{Beylich:2011aq}. 
 Further corrections to Eq.~(\ref{eq:benefit}) arise in extensions of  the SM with right-handed currents, which induce a finite  $A_T^{(2)}$ at low $q^2$. 
On the other hand, if SM-like chiralities are present only,
$A_T^{(2)}$ vanishes up to tiny corrections.
The latter  feature is in agreement with present data  \cite{Aaltonen:2011ja,LHCbMoriond2012}, although within sizeable uncertainties.
 Improved data and global analyses \cite{Altmannshofer:2011gn,Bobeth:2011nj} including checks of the OPE  \cite{Bobeth:2010wg} will exhibit whether such effects, which we neglect here, are contributing to  $\Delta B=1$ transitions. 

{\bf Analytic  form factors.}
For an analytic ansatz for the $q^2$-shape of the form factors we
employ the Series Expansion (SE)~\cite{Arnesen:2005ez,Boyd:1994tt,Boyd:1997qw,Caprini:1997mu,Becher:2005bg,Bourrely:2008za,Bharucha:2010im}.
It  is based on an expansion  in $z(t)$, defined as 
\begin{equation}
\label{eq:zdef}
z(t)\equiv 
z(t,t_0)
= \frac{ \sqrt{t_+-t}- \sqrt{t_+-t_0}}{\sqrt{t_+-t}+ \sqrt{t_+-t_0}} \, ,
\end{equation}
and  $t$ being the analytic continuation of $q^2$ to the complex plane, with $|z| \leq 1$. Here,
$t_{\pm}=(m_B \pm m_{K^*})^2$
and $0 \leq t_0<t_+$ for which we use 
$ t_0  =  t_+ (1 - \sqrt{1 - t_-/t_+})$ \cite{Hill:2006ub,Bharucha:2010im}.

We define the reduced transversity amplitudes 
$\hat f_i \equiv f_i/{\cal{N}}$, $i=\perp,||,0$, where the $q^2$-dependent normalization factor
${\cal{N}}$ is given in Eq.~\eqref{eq:factordef}.
The SE can then be written as:
 \begin{equation}
\label{par:SE}
\hspace{-3mm}
  \hat f_i
(t) 
= \frac{(\sqrt{-
z(t,0)
})^m
(\sqrt{
z(t,t_-)
})^l }{B
(t) 
\, \phi_f(t)
} \, \sum_{k} \alpha_{i,k} \, z^k
(t)
 \,,
 \end{equation}
with
$l=1,0,0$ and $m=1,1,0$ for $i=\perp,\parallel,0$, respectively.
The Blaschke factor  $B(t)=z(t,m_{R}^2)$
parametrizes an off-shell pole associated with a meson with
mass $m_R$; there can be several such factors if
multiple resonances are present.
 The function 
 $\phi_f
(t)
$ is given by
 \begin{align}
\phi_f
(t)
&=\sqrt{\frac{\eta}{48\pi\chi_f(n)}}
\, \frac{(t-t_+)}{(t_+-t_0)^{1/4}}
\left(\frac{
z(t,0)
}{-t}\right)^{(3+n)/2}
\nonumber\\
&\quad \times
\left(\frac{
z(t,t_0)
}{t_0-t}\right)^{-1/2}
\left(\frac{
z(t,t_-)
}{t_--t}\right)^{-3/4}\,,
\end{align}
with isospin factor $\eta=2$, and Wilson coefficients $\chi_f(n)$ with $n$ being the number of subtractions in the renormalization procedure (here $n=2$), discussed in detail in~\cite{Bharucha:2010im,Hill:2006ub}. We take $\chi_f(2)=1.2/(100 m_b^2)$. 
Note, that the ratios of the form factors, which are accessible with our analysis, are not sensitive  to this input.

\begin{table}[t]
\centering
\begin{tabular}{cc
rrrr} 
\hline
\hline
& BaBar   &
\multicolumn{2}{c}{CDF} 
&
\multicolumn{2}{c}{LHCb}
\\
$q^2$ [GeV$^2$]
& $F_L$ &
\multicolumn{1}{c}{$F_L$} & \multicolumn{1}{c}{$A_T^{(2)}$} & \multicolumn{1}{c}{$F_L$}  & \multicolumn{1}{c}{$A_T^{(2) \dagger}$}
\\
\hline
$[14.18,16]$ 
& $0.43^{+0.13}_{-0.16}$
& $0.29^{+0.15}_{-0.14}$&   $0.2^{+0.8}_{-0.8}$ 
& $0.35^{+0.10}_{-0.06}$ & $0.06^{+0.24}_{-0.29}$ 
\\
$[16,X]$  &  $0.55^{+0.15}_{-0.17}$
&  $0.20^{+0.20}_{-0.18}$  & $-0.7^{+0.9}_{-0.9}$  
& $0.37^{+0.07}_{-0.08}$ & $-0.75^{+0.35}_{-0.20}$
\\
\hline
\hline
\end{tabular}
\caption{
	Recent high-$q^2$ data from BaBar~\cite{BaBarLakeLouise2012}, CDF \cite{Aaltonen:2011ja} and LHCb~\cite{LHCbMoriond2012} with the statistical and systematic uncertainties added in quadrature. The maximum $q^2$-value in units of GeV$^2$
equals $X=19$ for LHCb and is the endpoint otherwise.$^\dagger$Using $A_T^{(2)}=2 S_3/ (1-F_L)$.
	\label{tab:data}}
\vspace{-2mm}
\end{table}
We are constricted to 4 d.o.f., {\it i.e.,} two bins in the high-$q^2$ region for each observable, see Table \ref{tab:data}.
Fortunately, we obtain a good fit already for  the lowest order SE, where
\begin{align}
	f_\perp
(t)
 &= \alpha_\perp\,
	\Lambda(t,m_{1^-}^2)
\sqrt{-
z(t,0)
} \sqrt{
z(t,t_-)
}
\,,
\cr 
f_\parallel
%(t)
 &=\alpha_\parallel\,
\Lambda(t,m_{1^+}^2)
\sqrt{-
z(t,0)
}
\,,
\cr  
f_0
(t) 
&= \alpha_0\,
\Lambda(t,m_{1^+}^2)
,
\label{eq:BV012par}
\end{align}
with
\begin{equation}
	\Lambda
(t,m_R^2)
=
\frac{{\cal{N} }}{
z(t,m_{\rm R}^2)
\,\phi_T^{V-A}
(t)
}\,,\;
\quad
\alpha_i \equiv \alpha_{i,0}\,.
\end{equation}
We take $m_{1^-}=5.42$~GeV for the  vector ($\perp$) and  $m_{1^+}=5.83$~GeV for the axial vector ($\parallel, 0$)  transitions~\cite{Nakamura:2010zzi}. 

\begin{figure}[t]
\centering
\includegraphics[width=0.23\textwidth]{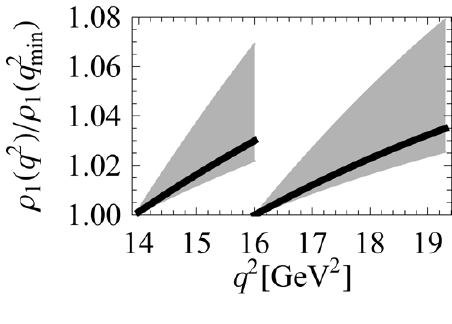}
\raisebox{0.13cm}{
\includegraphics[width=0.22\textwidth]{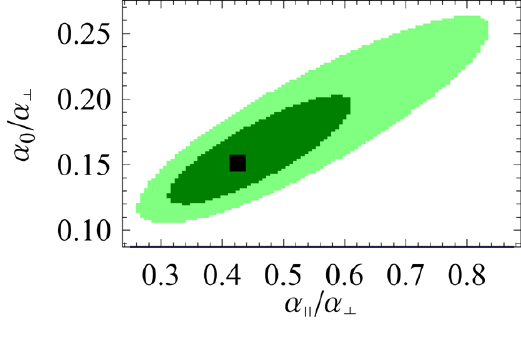}
}
\vspace{-3mm}
\caption{
	Left: The function 
	$\rho_1
(q^2)
$  normalized to its value at $q^2_{\rm min}=14$ and $16$~GeV$^2$ in the SM (solid curve) and with non-SM effects (shaded region).
	Right:
	Error correlation of $\alpha_\parallel/\alpha_\perp$ {\it vs.} $\alpha_0/\alpha_\perp$. The shown regions are $68\%$~CL (dark-green), $95\%$~CL (light-green) with the best-fit values
given by the black square.
}
\label{fig:rhoanderror}
\end{figure}

{\bf Fit and results.}
Even though Eqs.~\eqref{eq:fitform} is exact for each $q^2$, the data on the other hand is binned. 
\begin{figure}[t]
\centering
\includegraphics[width=0.36\textwidth]{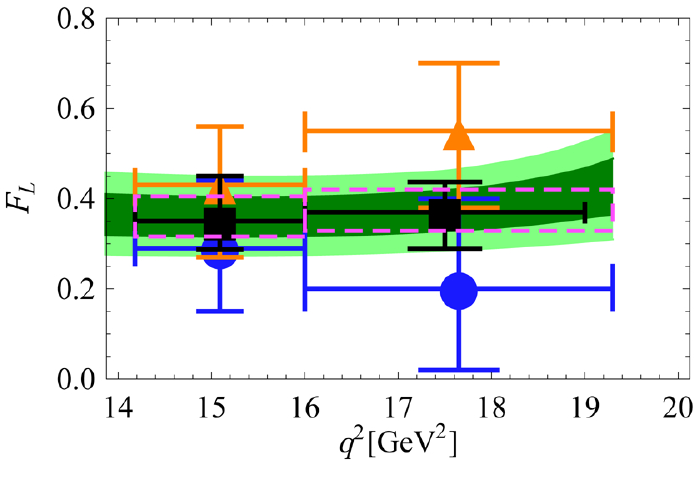}
\includegraphics[width=0.386\textwidth]{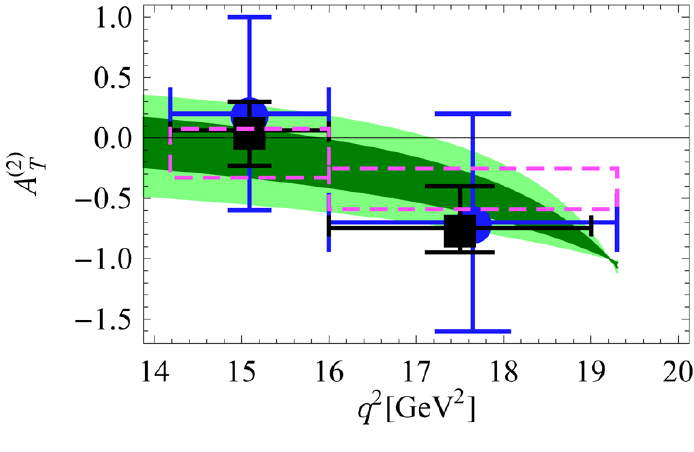}\;\;\;\;
\vspace{-1mm}
\caption{ 
Experimental data  by  BaBar~\cite{BaBarLakeLouise2012} (orange triangles), CDF~\cite{Aaltonen:2011ja} (blue circles) and LHCb~\cite{LHCbMoriond2012} (black squares), see Table \ref{tab:data}, versus  the outcome of the fit at 68$\%$~CL (dark-green bands unbinned, dashed magenta box binned) 95$\%$~CL (light-green bands unbinned) for $F_L$ (top) and $A_T^{(2)}$ (bottom). }
\vspace{-1mm}
\label{fig:expdataandfit}
\end{figure}
In the fit this is taken into account by replacing
$f_i^2(q^2)
$ with $ \int_{\textnormal{bin}} {\rm d}\, q^2 \,(2\,\rho_1\,(q^2)
f_i^2(q^2)
)
$, where the universal function $\rho_1 =(|C^L|^2+|C^R|^2)/2$ drops out of Eqs.~\eqref{eq:fitform} in the limit of vanishing bin-size. To the accuracy discussed after Eq.~(\ref{eq:ffbasis2}) we have
\begin{align}
	\rho_1
(q^2)
&=
\left|
\mathcal C_9^{\textnormal{eff}}
(q^2)
+
\kappa
\frac{2  m_b m_B}{q^2}
\mathcal C_7^{\textnormal{eff}}
(q^2)
\right|^2
+\left|
\mathcal C_{10}
\right|^2
\,,
\end{align}
where
$ \kappa=
1- 2 \frac{\alpha_s}{3 \pi} \ln \left(\mu/m_b \right) \simeq1$,
and 
$C_{j}^{\rm eff}=C_j+ ...$ denote the effective coefficients, see~\cite{Bobeth:2010wg} for details.
Fig.~\ref{fig:rhoanderror} (left-hand plot) shows the slope of $\rho_1$ in the SM
including next-to-leading order  QCD corrections as used in the fit (solid curve) and with physics beyond the SM (shaded areas). The latter is estimated
 by a variation of the Wilson coefficients
 $C_7$ from -0.4 to -0.3 and $| \mathcal C_{10}|$ from  2 to 5, see \cite{Bobeth:2011nj}.
 Since the bin-averaged shift in the slope is at the percent-level,  the uncertainties related to 
potential contributions from physics beyond the SM
are negligible given the current accuracy of the data.

\begin{figure*}[t]
\centering
\includegraphics[width=0.4\textwidth]{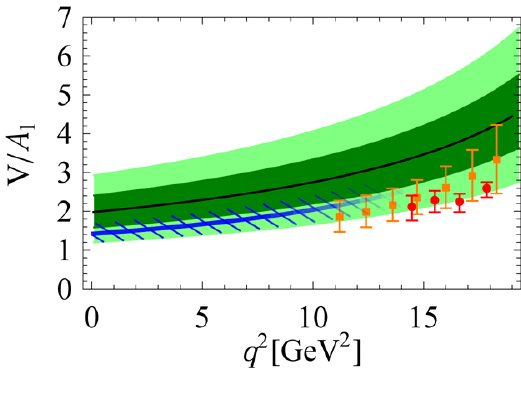}
\includegraphics[width=0.415\textwidth]{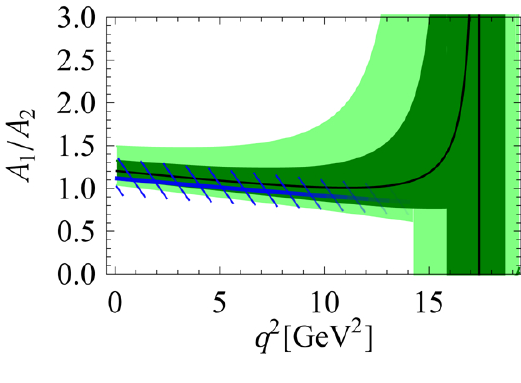}
\vspace{-3mm}
\\
\includegraphics[width=0.3\textwidth]{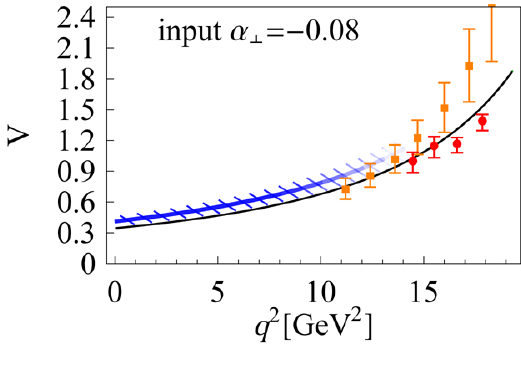}
\includegraphics[width=0.3\textwidth]{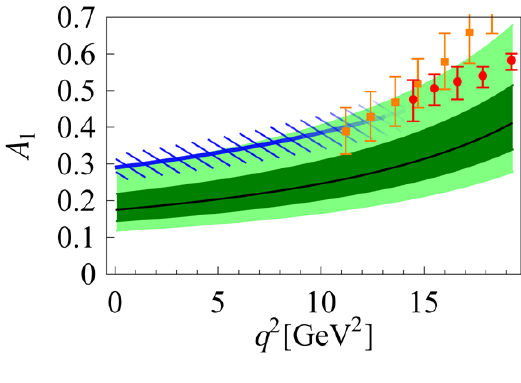}
\includegraphics[width=0.3\textwidth]{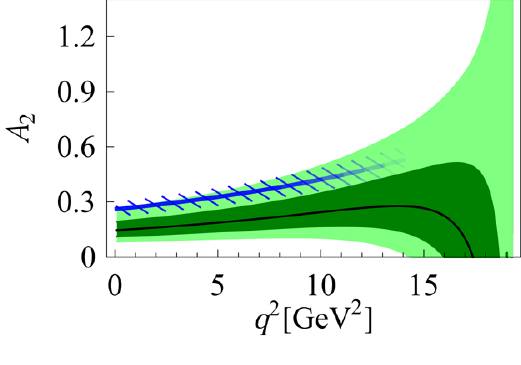}
\vspace{-3mm}
\caption{
	The form factor ratios $V/A_1$ and $A_1/A_2$ from fit to  data  \cite{Aaltonen:2011ja,BaBarLakeLouise2012,LHCbMoriond2012} (upper row). In the lower row the form factors are shown  using $V$ as  input with	$\alpha_\perp=-0.08$.
	The best fit result with  $68\%$~CL ($95\%$~CL) bands are given by the solid black line and the shaded dark-green (light-green) region, respectively.
The solid blue line and the hatched region denote LCSR results~\cite{Ball:2004rg}.
Lattice results from \cite{Liu:2011raa} and \cite{Becirevic:2006nm}, using the Isgur-Wise relations Eq.~(\ref{eq:IWR}), are shown as red circles and orange boxes, respectively. 
}
\label{fig:formfactors}
\end{figure*}

The SE in Eq.~\eqref{eq:BV012par} is truncated after the first order term, introducing three parameters $\alpha_\perp, \alpha_\parallel$ and $\alpha_0$. 
Since the data constrains only ratios of the $f_i^2$, we are sensitive to the modulus of the ratios of the $\alpha_i$ only. We choose to fit 
$\alpha_\parallel/\alpha_\perp$ and $\alpha_0/\alpha_\perp$ and obtain  $\chi_{\rm min}^2=3.8$ for all data, see Table~\ref{tab:data}.
The error correlation  of the fit is shown in Fig.~\ref{fig:rhoanderror} (right-hand plot), the fit to the data can be seen in Fig.~\ref{fig:expdataandfit}. 
The best-fit results with $1 \sigma$ errors are
\begin{equation}
	\hspace{-3mm} 
	\alpha_\parallel/\alpha_\perp=0.43^{+0.11}_{-0.08},
~
\alpha_0/\alpha_\perp=0.15^{+0.03}_{-0.02} ~~~~(\mbox{SE}).
 \end{equation} 
 The  resulting form factor ratios $V/A_1$ and $A_1/A_2$ are presented in 
 Fig.~\ref{fig:formfactors}.  We further show the form factors themselves,  fixing  $\alpha_\perp=-0.08$, {\it i.e.,} $V$, to allow for a comparison of the $q^2$-shapes.
As can be seen, the sensitivity to $A_2$ is very low towards the endpoint, caused by the
$\hat{\lambda}$ factor in the transversity amplitude $f_0$, see Eq.~(\ref{eq:ffdef}).
Also shown are  LCSR predictions for $q^2\lesssim 14$~GeV$^2$, where we
assigned flat uncertainties as obtained at $q^2=0$, {\it i.e.}, 11\%, 12\% and 14\% for $V,A_1$ and $A_2$, respectively~\cite{Ball:2004rg}.
Quenched \cite{Becirevic:2006nm} (set 3)  and unquenched lattice results~\cite{Liu:2011raa} are shown as well after  using the improved Isgur-Wise-relations to lowest order in $1/m_b$ \cite{Grinstein:2004vb,Bobeth:2010wg}
\begin{align} \label{eq:IWR}
T_1(q^2) = \kappa\, V\,(q^2) , \qquad T_2(q^2) = \kappa\, A_1 (q^2) .
\end{align}
We add theory uncertainties in quadrature, thereby ignoring possible cancellations  of systematics in the ratios.
Both lattice and LCSR predictions agree within 95\%CL with the ratios obtained from the SE fit.

Using the parametrization of the Simplified Series Expansion (SSE)~\cite{Bourrely:2008za}, corresponding to the changes
$B(t) \to 1-t/m_R^2$, $\phi_f(t) \to 1$, $\sqrt{-z(t,0)} \to \sqrt{t}/m_B$,  
$\sqrt{z(t,t_-)} \to \sqrt{\hat \lambda}$ in Eq.~(\ref{eq:BV012par}), a similarly good fit ist obtained with $\chi_{\rm min}^2=3.4$ and the best-fit results
\begin{equation}
	\hspace{-3mm}	\alpha_\parallel/\alpha_\perp=
0.44^{+0.11}_{-0.07},
~
\alpha_0/\alpha_\perp=
0.32^{+0.06}_{-0.04} ~~~~(\mbox{SSE}).
 \end{equation} 
The form factors from the SSE are in good agreement with the SE ones and not shown.
Noticeable differences arise only in the extrapolation to small $q^2$, where
the SSE results for $V/A_1$ can be about 50\% larger than the ones from SE, however
the predictions still overlap within $1 \sigma$.

Form factor ratios at $q^2=0$ along with LCSR predictions \cite{Ball:2004rg,Khodjamirian:2010vf} are compiled
in Table \ref{tab:ratios}.  Both fits return values of $V/A_1$ larger than the LCSR ones and larger than the symmetry-based prediction in the large energy limit
$V/A_1 =(m_B+m_{K^*})^2/(m_B^2+m_{K^*}^2-q^2)$~\cite{Charles:1998dr}, {\it i.e.,}$V(0)/A_1(0) = 1.3$  up to $1/m_b$ corrections~\cite{Burdman:2000ku}. 
A finer  binning  is expected to improve the control of the $q^2$-shape and
the performance of the fit results at low $q^2$. We stress, however, that already with present data a consistent picture over the whole physical range is obtained.

\begin{table}[t]
\centering
\begin{tabular}{ccccc} 
\hline
\hline
\mbox{}
& 
SE & SSE & LCSR$^*$~\cite{Ball:2004rg} & LCSR$^*$~\cite{Khodjamirian:2010vf}
\\
\hline
$V(0)/A_1(0)$ 
& $2.0 \pm 0.4$&   
$3.1 ^{+0.6}_{-0.7}$ 
& 
$1.4 \pm 0.2$ & 
$1.5 \pm 0.9$
\\
$A_1(0)/A_2(0)$  & 
$1.2 \pm 0.1$ &
$1.1 \pm 0.1 $ & 
$1.1\pm 0.2$ &
$1.0 \pm 0.7$
\\
\hline
\hline
\end{tabular}
\caption{Form factor ratios at 
$q^2=0$. $^*$Errors symmetrized and added in quadrature.
	\label{tab:ratios}}
\end{table}
{\bf Conclusions.}
Determinations of $B \to K^*$ form factor ratios from present data and theoretical predictions are 
mutually consistent, see Fig. \ref{fig:formfactors}.  
The currently still early  data may shift towards a larger  $A_T^{(2)}$ at low recoil and  lead to a correspondingly lower value of $V/A_1$, improving the agreement with the preliminary unquenched lattice determination. 
However, once data have become sufficiently precise allowing for consistency checks 
 \cite{,Bobeth:2010wg,Beylich:2011aq,Altmannshofer:2011gn,Bobeth:2011nj},
the low recoil predictions
 constitute a powerful test for  the lattice and its advances \cite{whitepaper}.

The method of form factor extractions from $B$-decays into a fully reconstructable final state
will greatly benefit  in the near-term future from the high-luminosity flavor searches at the LHC;
it is applicable to  other decays including
$\bar B_s\to \phi\; \mu^+ \mu^-$.
Both better statistics and  the assessment of further short-distance free 
observables \cite{Bobeth:2010wg} from a full angular analysis \cite{Kruger:1999xa} will 
improve the control of the low recoil region and of the form factors.

%
%
%
 
%%%%%%%%%%%%
{\bf Acknowledgments.}
We are happy to thank Aoife Bharucha, Christoph Bobeth,
Danny van Dyk,
Stefan Schacht and Matthew Wingate for useful exchanges.
This work is supported in part by the Bundesministerium f\"ur Bildung und Forschung (BMBF).

\vspace{-0.7cm}

%%%%%%%%%%%%%%%%%%%%%


\begin{thebibliography}{01}
\vspace*{3mm}


%\cite{Ball:1998kk}
\bibitem{Ball:1998kk} 
  P.~Ball and V.~M.~Braun,
  %``Exclusive semileptonic and rare B meson decays in QCD,''
  Phys.\ Rev.\ D {\bf 58}, 094016 (1998)
  [hep-ph/9805422].
  %%CITATION = HEP-PH/9805422;%%

%\cite{Liu:2011raa}
\bibitem{Liu:2011raa}
  Z.~Liu {\it at al.},
  %, S.~Meinel, A.~Hart, R.~R.~Horgan, E.~H.~Muller and M.~Wingate,
  %``A lattice calculation of B -> K(*) form factors,''
  arXiv:1101.2726 [hep-ph].
  %%CITATION = ARXIV:1101.2726;%%
%\cite{Bobeth:2010wg}
\bibitem{Bobeth:2010wg} 
  C.~Bobeth, G.~Hiller and D.~van Dyk,
  %``The Benefits of $\bar{B} -> \bar{K}^* l^+ l^-$ Decays at Low Recoil,''
  JHEP {\bf 1007}, 098 (2010)
  [arXiv:1006.5013 [hep-ph]].
  %%CITATION = ARXIV:1006.5013;%%


  
  %\cite{Kruger:2005ep}
\bibitem{Kruger:2005ep} 
  F.~Kruger and J.~Matias,
  %``Probing new physics via the transverse amplitudes of B0 ---> K*0 (---> K- pi+) l+l- at large recoil,''
  Phys.\ Rev.\ D {\bf 71}, 094009 (2005)
  [hep-ph/0502060].
  %%CITATION = HEP-PH/0502060;%%

%\cite{Aaltonen:2011ja}
\bibitem{Aaltonen:2011ja} 
  T.~Aaltonen {\it et al.}  [CDF Collaboration],
  %``Measurements of the Angular Distributions in the Decays $B \to K^{(*)} \mu^+ \mu^-$ at CDF,''
  Phys.\ Rev.\ Lett.\  {\bf 108}, 081807 (2012)
  [arXiv:1108.0695 [hep-ex]].
  %%CITATION = ARXIV:1108.0695;%%
  
  \bibitem{BaBarLakeLouise2012}
  S.~Akar for the BaBar Collaboration at the Lake Louise Winter Institute, Canada, February 23, 2012.
  
  \bibitem{LHCbMoriond2012} 
LHCb Collaboration,  
%	 ``Differential branching fraction and angular analysis of the $B^0 \to K^{* 0} \mu^+ \mu^- $ %decay,'' 
	 CERN-LHCb-CONF-2012-008.


  %\cite{Grinstein:2004vb}
\bibitem{Grinstein:2004vb} 
  B.~Grinstein and D.~Pirjol,
  %``Exclusive rare B [*RIGHTWARDS ARROW*] K*** [*SCRIPT SMALL L*] + [*SCRIPT SMALL L*] - decays at low recoil: Controlling the long-distance effects,''
  Phys.\ Rev.\ D {\bf 70}, 114005 (2004)
  [hep-ph/0404250].
  %%CITATION = HEP-PH/0404250;%%
  %\cite{Ball:2004rg}
\bibitem{Ball:2004rg}
  P.~Ball and R.~Zwicky,
  %``B_{d,s}->rho, omega, K*, phi Decay Form Factors from Light-Cone Sum Rules
  %Revisited,''
  Phys.\ Rev.\  D {\bf 71}, 014029 (2005)
  [arXiv:hep-ph/0412079].
  %%CITATION = PHRVA,D71,014029;%%
  
  %\cite{Beylich:2011aq}
\bibitem{Beylich:2011aq} 
  M.~Beylich, G.~Buchalla and T.~Feldmann,
  %``Theory of B -> K(*)l+l- decays at high q^2: OPE and quark-hadron duality,''
  Eur.\ Phys.\ J.\ C {\bf 71}, 1635 (2011)
  [arXiv:1101.5118 [hep-ph]].
  %%CITATION = ARXIV:1101.5118;%%

%\cite{Altmannshofer:2011gn}
\bibitem{Altmannshofer:2011gn} 
  W.~Altmannshofer, P.~Paradisi and D.~M.~Straub,
  %``Model-Independent Constraints on New Physics in b --> s Transitions,''
  arXiv:1111.1257 [hep-ph].
  %%CITATION = ARXIV:1111.1257;%%


%\cite{Bobeth:2011nj}
\bibitem{Bobeth:2011nj} 
  C.~Bobeth, G.~Hiller, D.~van Dyk and C.~Wacker,
  %``The Decay B --> K l^+ l^- at Low Hadronic Recoil and Model-Independent Delta B = 1 Constraints,''
  JHEP {\bf 1201}, 107 (2012)
  [arXiv:1111.2558 [hep-ph]].
  %%CITATION = ARXIV:1111.2558;%%
  


%\cite{Arnesen:2005ez}
\bibitem{Arnesen:2005ez}
  M.~C.~Arnesen, B.~Grinstein, I.~Z.~Rothstein and I.~W.~Stewart,
  %``A precision model independent determination of |V(ub)| from B --> pi e
  %nu,''
  Phys.\ Rev.\ Lett.\  {\bf 95}, 071802 (2005)
  [arXiv:hep-ph/0504209].
  %%CITATION = PRLTA,95,071802;%%
%\cite{Boyd:1994tt}
\bibitem{Boyd:1994tt}
  C.~G.~Boyd, B.~Grinstein and R.~F.~Lebed,
  %``Constraints On Form-Factors For Exclusive Semileptonic Heavy To Light Meson
  %Decays,''
  Phys.\ Rev.\ Lett.\  {\bf 74}, 4603 (1995)
  [arXiv:hep-ph/9412324].
  %%CITATION = PRLTA,74,4603;%%
%\cite{Boyd:1997qw}
\bibitem{Boyd:1997qw}
  C.~G.~Boyd and M.~J.~Savage,
  %``Analyticity, shapes of semileptonic form factors, and  anti-B --> pi l
  %anti-nu,''
  Phys.\ Rev.\  D {\bf 56}, 303 (1997)
  [arXiv:hep-ph/9702300].
  %%CITATION = PHRVA,D56,303;%%
%\cite{Dalgic:2006dt}
%\cite{Caprini:1997mu}
\bibitem{Caprini:1997mu}
  I.~Caprini, L.~Lellouch and M.~Neubert,
  %``Dispersive bounds on the shape of anti-B --> D(*) l anti-nu form
  %factors,''
  Nucl.\ Phys.\  B {\bf 530}, 153 (1998)
  [arXiv:hep-ph/9712417].
  %%CITATION = NUPHA,B530,153;%%
%\cite{Becher:2005bg}
\bibitem{Becher:2005bg}
  T.~Becher and R.~J.~Hill,
  %``Comment on form factor shape and extraction of |V(ub)| from B --> pi l
  %nu,''
  Phys.\ Lett.\  B {\bf 633}, 61 (2006)
  [arXiv:hep-ph/0509090].
  %%CITATION = PHLTA,B633,61;%%



%\cite{Bourrely:2008za}
\bibitem{Bourrely:2008za}
  C.~Bourrely, I.~Caprini and L.~Lellouch,
  %``Model-independent description of $B\to \pi l\nu$ decays and a determination
  %of $|V_{ub}|$,''
  Phys.\ Rev.\  D {\bf 79}, 013008 (2009)
  [Erratum-ibid.\  D {\bf 82}, 099902 (2010)]
  [arXiv:0807.2722 [hep-ph]].
  %%CITATION = PHRVA,D79,013008;%%

%\cite{Bharucha:2010im}
\bibitem{Bharucha:2010im} 
  A.~Bharucha, T.~Feldmann and M.~Wick,
  %``Theoretical and Phenomenological Constraints on Form Factors for Radiative and Semi-Leptonic B-Meson Decays,''
  JHEP {\bf 1009}, 090 (2010)
  [arXiv:1004.3249 [hep-ph]].
  %%CITATION = ARXIV:1004.3249;%%


%\cite{Hill:2006ub}
\bibitem{Hill:2006ub}
  R.~J.~Hill,
  %``The modern description of semileptonic meson form factors,''
%{\it In the Proceedings of 4th Flavor Physics and CP Violation Conference (FPCP 2006), %Vancouver, British Columbia, Canada, 9-12 Apr 2006, pp
%027}
  [arXiv:hep-ph/0606023].
  %%CITATION = ECONF,C060409,027;%%


%\cite{Nakamura:2010zzi}
\bibitem{Nakamura:2010zzi} 
  K.~Nakamura {\it et al.}  [Particle Data Group Collaboration],
  %``Review of particle physics,''
  J.\ Phys.\ G G {\bf 37}, 075021 (2010).
  %%CITATION = JPHGB,G37,075021;%%

%\cite{Becirevic:2006nm}
\bibitem{Becirevic:2006nm} 
  D.~Becirevic, V.~Lubicz and F.~Mescia,
  %``An Estimate of the B ---> K* gamma form factor,''
  Nucl.\ Phys.\ B {\bf 769}, 31 (2007)
  [hep-ph/0611295].
  %%CITATION = HEP-PH/0611295;%%

%\cite{Khodjamirian:2010vf}
\bibitem{Khodjamirian:2010vf} 
  A.~Khodjamirian, T.~Mannel, A.~Pivovarov and Y.~Wang,
  %``Charm-loop effect in $B \to K^{(*)} \ell^{+} \ell^{-}$ and $B\to K^*\gamma$,''
  JHEP {\bf 1009}, 089 (2010)
  [arXiv:1006.4945 [hep-ph]].
  %%CITATION = ARXIV:1006.4945;%%
  
  %\cite{Charles:1998dr}
\bibitem{Charles:1998dr} 
  J.~Charles {\it et al.},
  %, A.~Le Yaouanc, L.~Oliver, O.~Pene and J.~C.~Raynal,
  %``Heavy to light form-factors in the heavy mass to large energy limit of QCD,''
  Phys.\ Rev.\ D {\bf 60}, 014001 (1999)
  [hep-ph/9812358].
  %%CITATION = HEP-PH/9812358;%%
  
  %\cite{Burdman:2000ku}
\bibitem{Burdman:2000ku} 
  G.~Burdman and G.~Hiller,
  %``Semileptonic form-factors from B ---> K* gamma decays in the large energy limit,''
  Phys.\ Rev.\ D {\bf 63}, 113008 (2001)
  [hep-ph/0011266].
  %%CITATION = HEP-PH/0011266;%%

\bibitem{whitepaper}
USQCD collaboration, Whitepaper "Lattice QCD and High-Intensity Flavor Physics" November 11, 2011.


%\cite{Kruger:1999xa}
\bibitem{Kruger:1999xa} 
  F.~Kruger, L.~M.~Sehgal, N.~Sinha and R.~Sinha,
  %``Angular distribution and CP asymmetries in the decays anti-B ---> K- pi+ e- e+ and anti-B ---> pi- pi+ e- e+,''
  Phys.\ Rev.\ D {\bf 61}, 114028 (2000)
  [Erratum-ibid.\ D {\bf 63}, 019901 (2001)]
  [hep-ph/9907386].
  %%CITATION = HEP-PH/9907386;%%
  
  


\end{thebibliography}
\end{document}